\documentclass{aa}  
\usepackage{graphicx}
\usepackage{txfonts}
\usepackage{natbib}

\usepackage{ulem}
\include{definitions}
      \def\new#1 {{\bf #1 }}
      \def\cut#1 {\sout{#1} }

\include{definitions}
\begin{document}
\title{Searching for coronal radio emission from protostars using Very-Long-Baseline Interferometry}
\author{J. Forbrich\thanks{now at: Harvard-Smithsonian Center for Astrophysics, 60 Garden Street, MS 42, Cambridge, MA 02138, U.S.A.,  \email{jforbrich@cfa.harvard.edu}} \and M. Massi \and E. Ros \and A. Brunthaler \and K.~M. Menten}
\institute{
Max-Planck-Institut f\"ur Radioastronomie, Auf dem H\"ugel 69, D-53121 Bonn, Germany
}
 
\date{Received; accepted}

\abstract{}{In order to directly study the role of magnetic fields in the immediate vicinity of protostars, we use Very-Long-Baseline Interferometry (VLBI), aiming at the detection of non-thermal centimetric radio emission. This is technically the only possibility to study coronal emission at sub-AU resolution.}{We performed VLBI observations of the four nearby protostars HL~Tau, LDN~1551~IRS5, EC~95, and YLW~15 in order to look for compact non-thermal centimetric radio emission. For maximum sensitivity, we used the High Sensitivity Array (HSA) where possible, involving the Very Long Baseline Array (VLBA), the phased Very Large Array (VLA), as well as the Arecibo, Green Bank, and Effelsberg radio telescopes.}{While all four protostars were detected in VLA-only data, only one source (YLW\,15\,VLA\,2) was detected in the VLBI data. The possibility of non-detections due to free-free absorption, possibly depending on source geometry, is considered. For YLW\,15\,VLA\,2, the prospects for an accurate orbit determination appear to be good.}{}


\maketitle

\section{Introduction}

According to the classification scheme suggested by \citet{lad87}, low-mass pre-main sequence (PMS) stars form an evolutionary sequence. The sequence starts with the youngest (age $\lesssim10^4$ y) ``class~0'' sources \citep{and93}, that are not detected at near-infrared wavelengths and are deeply embedded in very dense dust envelopes of low temperature. It then proceeds over less embedded class~I sources (age $10^5$ y), to virtually envelope-free, optically visible, class~II protostars (classical T Tau stars, $10^6-10^7$ y). After being class~III protostars or weak-lined or naked T~Tau stars (i.e., without or with only very little disk material), they reach the zero-age main sequence at ages $>10^7$ y. For a discussion of radio properties of Young Stellar Objects (YSOs), see, e.g., \citet{and96}.

Powerful X-ray  and gyrosynchrotron radio-continuum flares, orders of magnitudes
stronger than solar flares, indicate that protostars exhibit unusually high
levels of magnetic activity. \citet{fem99} identify four possible magnetic geometries in which hard X-ray and non-thermal radio emission
might arise:

\begin{itemize}
\item[(1)] Solar-type multipolar fields with both foot-prints rooted
in the stellar photosphere.
\item[(2)] Field lines connecting the star to the circumstellar disk
at its co-rotation radius. This could be the explanation for the quasi-periodic X-ray flares observed towards the class~I protostar YLW\,15 \citep{tsu00a,mon00}.
\item[(3)] Field lines above the co-rotation radius.
\item[(4)] Magnetic loops with  both origins in the disk.
\end{itemize}
   
\noindent The presence of a disk is fundamental  for magnetic energy release in the three last cases, but irrelevant for the first case. Further possible configurations include interacting coronal structures in binaries (see below). Apart from gyrosynchrotron emission, also completely different non-thermal emission mechanisms can occur, e.g., coherent emission like electron cyclotron maser radiation \citep{dul85,smi03}.

In the late 1980s, the first VLBI observations of YSOs were reported. \citet{fel89} described the VLBI detection of radio emission from the $\theta^1$\,Ori\,A multiple system; $\theta^1$\,Ori\,A is one of the
''Trapezium stars'' at the center of the young Orion Nebula Cluster. \citet{and91} reported non-thermal gyrosynchrotron radiation towards the young magnetic B star $\rho$~Oph~S1. \citet{and92} published a VLBI survey of the $\rho$~Oph star-forming region, showing that several T Tauri stars have non-thermal radio emission, although detections did not include embedded protostellar sources (like IRAS~16293--2422). The very active system HD~283447 (= V773\,Tau), located in the Taurus-Auriga molecular cloud complex at a distance of $d$=140\,pc, could be resolved at VLBI scales \citep{phi91,phi96}. It consists of two class~III objects. The VLBI observations have shown that the emission arises from regions of some tenths of an AU, 
and that the bulk of the radiation comes from two small unresolved components related to the two stars.
The reason of the large flares observed in this system is different from the mechanisms listed above; the flares are clustered around the periastron passage of the binary system and therefore probably due to colliding interbinary stellar coronal structure involving helmet streamers \citep{mas02,mas06}. As a matter of fact, \citet{phi91,phi96} have performed several other VLBI observations of this source outside the periastron flaring window and observed a source of  a size of a few stellar radii, or anyway much smaller than the dimensions of the binary system.

In order to study the predicted role of the disk as described by \citet{fem99}, it is perhaps most interesting to observe younger objects than class~III. This is difficult, since only very few candidates are known (i.e., protostars showing both X-ray and non-thermal radio emission). It is also possible that a putative weak radio source is embedded in an absorbing plasma showing thermal (free-free) emission. The magnetic fields which lead to magnetic reconnection and generate flares at small scales ($\lesssim 0.1$~AU) play an important role in driving thermal radio jets and bipolar outflows; the shock front produced by the collision of the outflows with the ISM ionizes part of the gas and creates ionized regions appearing elongated at VLA resolutions (e.g., \citealp{tsu04}). Any emission process taking place at  smaller scales will suffer from significant free-free absorption and will be difficult to detect. This is exactly the mechanism invoked by \citet{gir04} to explain their non-detection with the VLBA at 6\,cm of the class~I protostar YLW 15. 
\citet{and87} estimates how easily emission from within this region can be entirely concealed by ionized material: A mass-loss wind of only 10$^{-11}$~M$_\odot$\,yr$^{-1}$ suffices to entirely absorb 5~mJy of flux density of $T_b > 10^7$~K from behind.
In conclusion, the real problem in VLBI observations of protostars younger than class~III is that the weak non-thermal emission suffers from strong free-free absorption. Additionally, free-free emission observed at angular scales probed by the VLA may be ``resolved out'' at VLBI scales. The shortest baseline in the VLBI data discussed here corresponds to angular scales of about 0\farcs1. Thus, already emission of slightly larger extent remains undetected. However, the rare detections of non-thermal radio emission towards protostars may also reflect an intrinsic property of these sources.

The VLBI observations (using the VLBA, the VLA, and the Effelsberg 100-m telescope) at a wavelength of 3.6~cm by \citet{smi03} of the class~II protostar T Tau-S are very important in this respect: Weak and variable emission (between 1.4 and 3.8\,mJy) was finally detected, even if unresolved with the 1\,mas HPW beam. From multi-epoch VLBI observations of the same source, \citet{loi05} confirm the detection of a compact radio source and derive a parallax yielding a precise distance ($141.5_{-2.7}^{+2.8}$~pc).

In this paper, we report the results of our search for compact, coronal radio emission towards protostars, using high-sensitivity VLBI observations.
In Sect.~\ref{sect.targ}, we describe our target selection procedure before a description of the actual VLBI observations and their analysis in Sect.~\ref{sect.obse}. The results are described in Sect.~\ref{sect.resu} and finally, we summarize our findings in Sect.~\ref{sect.summ}.

\begin{table*}
\caption{Journal of observations$^1$}             
\label{jourobs}      
\centering                          
\begin{tabular}{lllllrrrl}        
\hline\hline                 
Source & $d$[pc]&Phase ref. & Project & Date & Time (UTC) & On-source$^2$ & VLA$^3$ & Comments \\    
\hline                        
YLW~15                &130& J1625-2517 & BF083  & 30 Apr 2005 & 06:15--10:15 & 142.0~min & B   & VLBA without LA$^4$\\
HL~Tau \& LDN\,1551$^5$ &140& J0431+1731 & BF084a & 03 Jul 2005 & 12:30--16:30 &  64.1~min & BnC & Arecibo data lost\\
EC~95                 &310& J1838+0404 & BF084b & 29 Jul 2005 & 01:00--04:00 &  90.4~min & C   & --\\
HL~Tau \& LDN\,1551$^5$ &140& J0431+1731 & BF089  & 24 May 2006 & 15:00--19:00 &  65.5~min & AnB & BF84a reobserved\\
\hline                                   

\multicolumn{7}{l}{$^1$ carried out with the High Sensitivity Array (HSA) except for BF83 (VLBA+VLA+GBT)} \\
\multicolumn{5}{l}{$^2$ scan minutes, not taking into account the number of antennas observing} \\
\multicolumn{5}{l}{$^3$ configuration of the VLA during the observation} \\
\multicolumn{5}{l}{$^4$ the Los Alamos VLBA antenna} \\
\multicolumn{5}{l}{$^5$ IRS~5} \\
\end{tabular}
\end{table*}

\section{Target selection}
\label{sect.targ}

In order to establish the role of the disk or star/disk interaction,
we planned to observe nearby protostars at an earlier stage than class~III at $\lambda=$ 3.6~cm using VLBI, trying to detect non-thermal radio emission. Given the difficulties described, we first had to define a good sample of target sources. We selected class~I and class~II protostellar sources which were detected in centimetric radio as well as in (variable) X-ray emission as tracers of coronal magnetic activity. Since information on polarized radio emission from protostars---a sign for non-thermal emission---is sparse, sources were selected that had either variable radio emission or a negative spectral index indicative of non-thermal radio emission. Only few sources are known to fulfil these criteria so that no additional criteria (e.g., distance) were applied. In our actual observations (see below), synthesized beam sizes of down to $0.2 \times 0.1$~AU were realized. We chose an observing frequency of 8.4~GHz ($\lambda=$ 3.6~cm) as a compromise between a reduced influence of free-free-absorption (compared to longer wavelengths) and ensuring a good sensitivity by using a standard frequency allowing the use of the Arecibo 305m telescope. 

Our sample consists of the class~0/I binary protostar YLW\,15, the class~I binary protostar LDN\,1551~IRS5, the class~II source HL~Tau, as well as the proto-Herbig Ae/Be star EC~95. Distances range from 130~pc to 310~pc (see Table~\ref{jourobs}). As noted above, YLW\,15 was observed with VLBI before \citep{gir04}. Except for this source (YLW\,15), all sources are in the declination range accessible to the Arecibo 305m telescope and thus were observed with the full High Sensitivity Array (HSA), consisting of the VLBA, the phased VLA, Arecibo, GBT, and the Effelsberg 100-m telescope. The radio luminosities of these sources, as deduced from previous VLA observations, are all very similar, as will become clear in the source descriptions.

\subsection{The binary protostar YLW\,15}

YLW\,15 is a binary protostar whose components are in the class~0/I stages. It is located in the $\rho$~Oph dark cloud at a distance of 130~pc \citep{reb04}, and it shows X-ray and radio emission. It was first found to be a radio binary (with components named VLA\,1 and VLA\,2, \citealp{gir00}), whose orbital parameters were estimated by \citet{cur03} from multi-epoch VLA observations.
After an early VLBI attempt to detect non-thermal radio emission from the YLW\,15 system yielded an upper flux density limit of 1\,mJy \citep{and92}, recent VLBI observations were reported by \citet{gir04} who observed at a wavelength of 6~cm and achieved an upper limit of 0.2~mJy (4$\sigma$). However, the authors note that both the non-detection of the source on the VLA-Goldstone baseline and their finding that the source has a non-variable radio spectrum point towards an extended source of thermal free-free emission. Interestingly, their additional near- and mid-infrared data indicates that VLA\,1 is more deeply embedded (or less luminous) than VLA\,2 while the orbital motion \citep{cur03} indicates that VLA\,1 is the more massive object.

VLA\,2 is responsible for strong quasi-periodic X-ray emission explained as due to the fast rotation of the star with respect to the disk \citep{tsu00a,mon00}. The consequent star-disk shearing of the magnetic field lines gives rise to  magnetic reconnection and therefore creation of energetic electrons producing the quasi-periodic X-ray emission. Non-thermal radio emission is therefore expected. The integrated flux at 3.6~cm, derived from VLA measurements, is S$_{3.6 \rm cm}=0.64\ldots0.78$~mJy \citep{gir04}. When comparing the VLBA-only non-detection at a wavelength of 6~cm with VLA data at the same wavelength, it seems that at least two thirds of the total emission is thermal emission from a collimated outflow which is indeed partially resolved at VLA scales \citep{gir04}. The non-thermal contribution is thus embedded in an absorbing thermal medium. 

\subsection{The binary protostar LDN 1551 IRS 5}

This source, located at a distance of 140~pc in Taurus, is one of the best known deeply embedded young stellar objects. It drives one of the most spectacular bipolar outflows, the first ever detected \citep{sne80}, extending over more than $10'$ (e.g., \citealp{mor88}). The dark cloud LDN~1551 itself is one of the nearest and most active regions of ongoing low-mass stars formation. IRS~5 is the most luminous YSO in this region ($L= 40$~L$_\odot$), and deeply embedded ($A_V \approx 150$~mag, e.g., \citealp{whi00}). \citet{rod98} discovered that IRS~5 is a protobinary separated by 40~AU (see also \citealp{rod86}). Based on VLA observations at $\lambda = 7$~mm, they detected two protoplanetary disks with semimajor axes of about 10~AU. The integrated flux densities at $\lambda = 3.6$~cm are 0.78~mJy and 0.70~mJy for the northern and southern component, respectively \citep{rod98}. Most recently, \citet{lim06} analyze the source at 7~mm using the VLA combined with the Pie Town VLBA antenna, and find that LDN\,1551~IRS\,5 is actually a triple system.

\citet{fav02} detected a faint X-ray source near IRS~5 in an observation carried out with XMM-\textsl{Newton}. \citet{bal03}, presenting \textsl{Chandra} X-ray data of higher angular resolution (but still not resolving the binary), conclude from a position offset and a difference in absorbing column density that an X-ray
source is located at the base of the HH~154 jet. The source, named ACIS~31, is located about $0\farcs5$ west of IRS~5 with 60 counts registered in the $0.5-8$~keV band in a 79~ksec exposure, 47 of them coming from within $1\farcs7$ of the source.  Due to the difference in angular resolution between the XMM-\textsl{Newton} and \textsl{Chandra} data, it cannot be excluded that the X-ray emission is instead due to the protostar itself (although that appears to be unlikely; see \citealp{fav06} who interpret morphological changes in the source as due to jet motion). 

\subsection{The class~I/II transition protostar HL~Tau}

HL~Tau is a flat (optical---far-infrared) spectrum T Tauri star, deeply embedded in a dusty nebulosity associated with a reflection nebula. In the evolutionary sequence, it is in a transition phase between class~I and class~II. There is strong extinction towards the central source ($A_V > 22$ mag, \citealp{sta95}), although the star is detected in the near--infrared \citep{wei95,bec95}. The morphology of the optical and NIR reflection nebula as well as the flat spectral energy distribution of HL~Tau can be explained by scattering and thermal emission from an infalling, dusty protostellar envelope typical of Taurus class~I sources \citep{cal94}. The star is associated with the outflow system HH~266. At centimeter radio wavelengths, HL~Tau has been analyzed by \citet{rod92,rod94}, using the VLA. At $\lambda =1.3$~cm, they observe a structure elongated approximately in the north-south direction surrounding the star (about 1.4 x 0.6 arcsec, or about 200 x 80 AU at a distance of 140~pc). The authors argue that the size and the integrated flux density of this source ($2.9 \pm 0.4$ mJy) at $\lambda =1.3$~cm indicate that the emission originates from dust in a protoplanetary disk. At $\lambda =3.6$~cm, the picture is strikingly different: Here, the observed structure is elongated approximately in the east-west direction (with an integrated flux density of $0.52 \pm 0.02$ mJy).
The authors argue that in contrast to the emission at $\lambda =1.3$~cm, the emission at $\lambda =3.6$ cm is dominated by free-free radiation from an ionized outflow. In order to study the structure of its circumstellar material, HL~Tau has been studied extensively at millimeter and submillimeter wavelengths \citep[][]{lay94,mun96,lay97,loo00}. 
\citet{fav03} discuss X-ray data of HL~Tau. While the source seems to have been quiescent during XMM-\textsl{Newton} observations, there are \textsl{Chandra} data containing evidence for a short-duration flare.

\subsection{The Proto-Herbig Ae/Be star EC 95}

EC~95 is a very young intermediate-mass star (i.e., up to 5 M$_\odot$), presumably a precursor of a Herbig AeBe (HAeBe) star, therefore not a low-mass protostar like our other sources. It is located in the Serpens cloud core at a distance of 310~pc.

The source has an X-ray luminosity of $3.7\times10^{31}$ erg\,s$^{-1}$ \citep{pre03}. Radio emission has been detected (with peak flux densities $S_\mathrm{8.4\,GHz}\sim0.6$\,mJy and $S_\mathrm{5\,GHz}\sim 0.8$\,mJy) by \citet{smi99}. The slightly negative spectral index (in $S \propto \nu^\alpha$) is inconsistent with thermal emission from an optically thick wind, where rather $\alpha=+2$ is expected \citep{sim83}. There is also evidence for variability on timescales of years. \citet{eir05} report a peak flux density of $1.65\pm0.1$~mJy on 24 Oct 1993, also from 3.5~cm observations carried out with the VLA in CnD configuration. Only about one month earlier (in the observations reported by \citealp{smi99}), the source was more than three times fainter.

Although intermediate-mass stars are thought to be fully radiative, and therefore should not have a corona, previous detections of X-rays from HAeBe stars have been made (e.g., \citealp{zip94}). \citet{tou95} suggested that radially differential rotation in intermediate-mass stars could lead to the formation of a temporary convective layer and hence a corona. They estimated that the lifetime of such a corona would be of order 10$^6$ years, approximately the timescale for intermediate-mass objects to evolve onto the main sequence.
On the other hand, a companion star responsible for the corona cannot be excluded. Another explanation of X-ray emission from Herbig Ae/Be stars involves star-disk interactions \citep{ham05b}.

\begin{figure*}
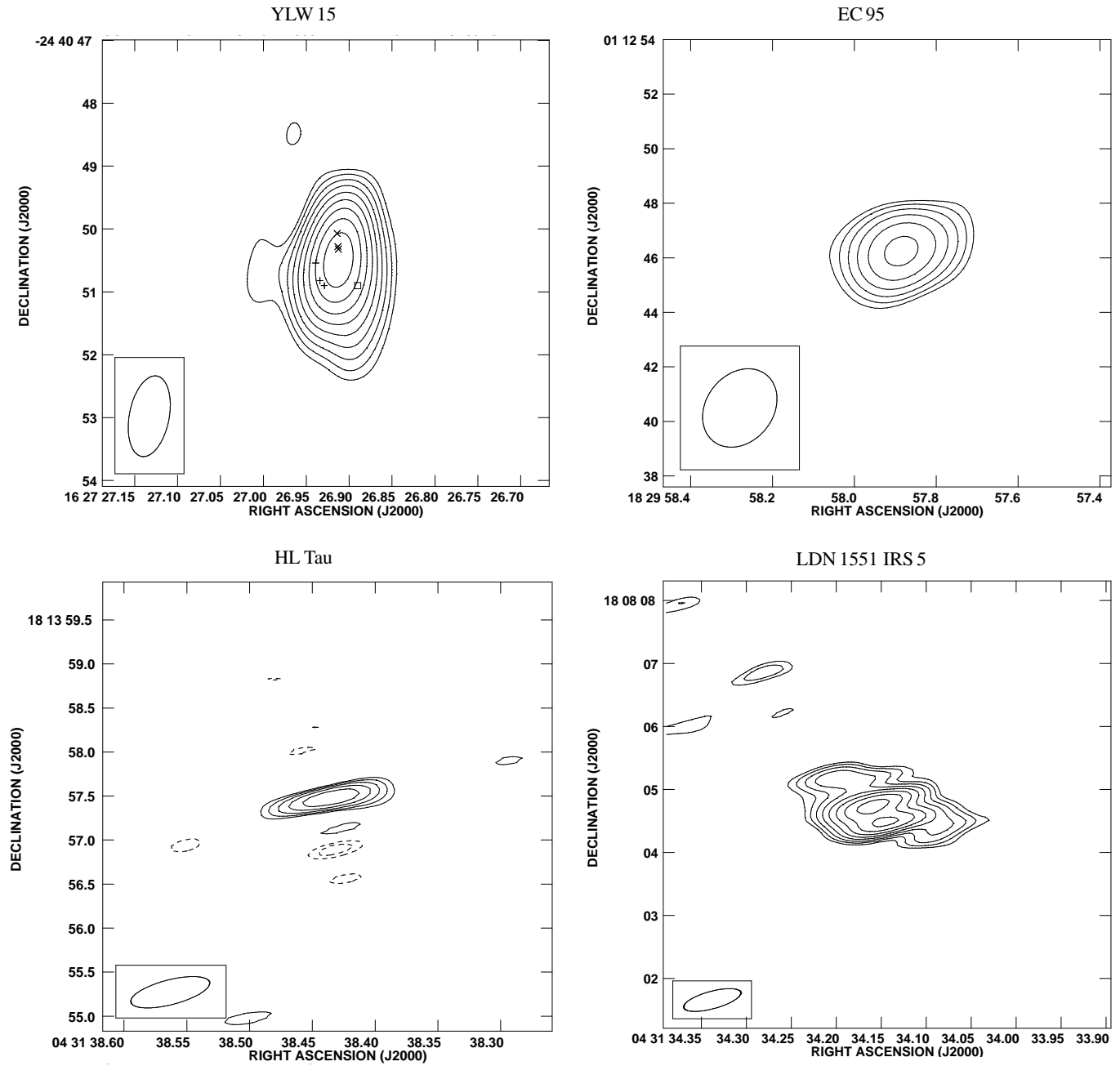

\begin{minipage}{9cm}
      \centerline{\hspace{7mm}YLW\,15}
      \vspace{2mm}
      \centerline{\hbox{
      \includegraphics*[width=8.9cm, bb=40 176 570 644]{BF083_YLW15_K3.PS}
      }}
\end{minipage}
\begin{minipage}{9cm}
      \centerline{\hspace{7mm}EC\,95}
      \vspace{2mm}
      \hspace{1mm}
      \centerline{\hbox{
      \includegraphics*[width=8.7cm, bb=45 173 570 647]{BF084B_EC95_VLAK3N.PS}
      }}
\end{minipage}

\vspace{4mm}

\begin{minipage}{9cm}
      \vspace{1mm}
      \centerline{\hspace{7mm}HL\,Tau}
      \vspace{2mm}
      \centerline{\hbox{
      \includegraphics*[width=9.0cm, bb=45 178 570 640]{BF089_HLTAU_VLAK3.PS}
      }}
\end{minipage}
\begin{minipage}{9cm}
      \centerline{\hspace{7mm}LDN\,1551~IRS\,5}
      \vspace{2mm}
      \hspace{1mm}
      \centerline{\hbox{
      \includegraphics*[width=8.7cm, bb=45 175 570 644]{BF089_L1551_VLAK3.PS}
      }}
\end{minipage}
\caption{VLA-only maps of the four program sources YLW\,15-VLA2, EC95, HL~Tau, and LDN\,1551~IRS5. Note that YLW\,15 is dominated by VLA\,1 (see also Fig.~\ref{vlaonlyYLW15} for an image restored with a circular beam smaller by about a factor of two). For the two components of YLW\,15, the proper motions as read from Fig.~2 in \citet{cur03} are shown (`$\times$' for VLA\,1 and '+' for VLA\,2; years 1990, 2000, and 2002 from north to south) together with the correlation position (`$\square$'), see text. For all sources the contour lines delineate multiples of the 3$\sigma$ image noise, $(-2,-\sqrt{2},-1,1,\sqrt{2}, 2,\ldots)\times3\sigma$, increasing by factors of $\sqrt{2}$. The respective $3\sigma$ levels are 72~$\mu$Jy (YLW~15),  79~$\mu$Jy (EC~95), 72~$\mu$Jy (HL~Tau), and 74~$\mu$Jy (LDN\,1551).}
\label{vlaonly}
\end{figure*}

\section{Observations and data analysis}
\label{sect.obse}

In order to reach the highest possible sensitivity we used the world's biggest radio telescopes together with the NRAO Very Long Baseline Array (VLBA, ten antennas of 25~m diameter each), most notably the so-called High Sensitivity Array (HSA) consisting of the Arecibo 305~m, the Effelsberg and Green Bank (GBT) 100-m radio telescopes, the phased Very Large Array (VLA, equivalent in collecting area to a 130~m dish) as well as the VLBA. The usage of the phased VLA, equivalent in sensitivity to a large single-dish telescope of 130~m diameter, has the additional advantage of providing regular VLA data as well. Thus, the source positions can be checked before correlation. All runs were phase-referencing observations in dual polarization (RCP \& LCP) taken at 256\,Mbit\,s$^{-1}$, eight baseband channels of 8~MHz bandwidth covering an aggregate bandwidth of 32~MHz, and 2-bit sampling (mode 4cm 256-8-2-UL).
Each observation was correlated with 16 channels per IF. The field of view is limited by time-smearing due to the correlator averaging time. Its size is estimated to be $\sim1\farcs8$ for YLW~15 and $\sim3\farcs3$ otherwise. Basic observation parameters including the phase reference sources used are listed in Table~\ref{jourobs}. The phase reference sources were at $52'$ (YLW15), $\sim40'$ (HL~Tau \& LDN\,1551~IRS5), and $3.6^\circ$ (EC~95) of the program sources. We spent about three minutes on the program source and 1.5 minutes on the reference source in every cycle.

\begin{figure}
  \centerline{YLW\,15 (VLA-only, circular $0\farcs4$ beam)}
  \vspace{2mm}
  \centerline{\hbox{
      \includegraphics*[width=8.5cm, bb=45 182 570 637]{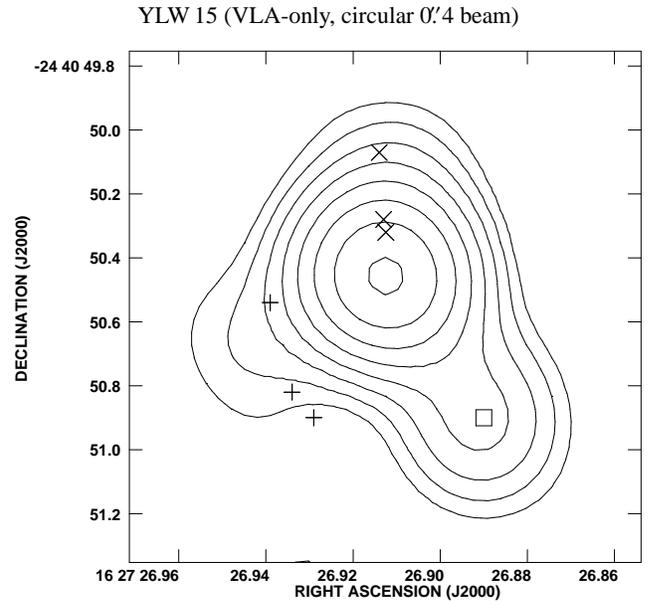}      }}
\caption{VLA-only map of YLW\,15, restored with a circular $0\farcs4$ beam. For the two components of YLW\,15, the proper motions as read from Fig.~2 in \citet{cur03} are shown (`$\times$' for VLA\,1 and '+' for VLA\,2; years 1990, 2000, and 2002 from north to south) together with the correlation position (`$\square$'). The contour lines delineate multiples of 0.1~mJy ($~\sim4\sigma$), increasing by factors of $\sqrt{2}$ (as explained for Fig.~\ref{vlaonly}).}
\label{vlaonlyYLW15}
\end{figure}

In order to obtain accurate positions for the correlation of the VLBI data, the VLA data for all sources were analyzed first (Fig.~\ref{vlaonlyYLW15}, see text below). Confirming the position was especially important in the case of YLW\,15 due to the apparent orbital motion of its components. All VLA and VLBI data were analyzed with the NRAO Astronomical Image Processing System (AIPS).
The VLA data of YLW~15 and EC~95 were amplitude-calibrated using a standard total flux density of 3C\,286 of 5.22~Jy while HL~Tau and LDN\,1551~IRS\,5 were calibrated using a standard total flux density of 3C\,84 of 3.17~Jy. While EC~95 and HL~Tau were correlated simply at the positions of the singular emission peaks, and LDN\,1551~IRS\,5 was correlated at the position of the northern component, YLW\,15-VLA2 was correlated at a position in the south-western extension to the main source as seen more clearly in a VLA-only image restored with a circular $0\farcs4$ beam (Fig.~\ref{vlaonlyYLW15}). The correlation parameters used are listed in Table~\ref{corpars} for all sources.

\begin{table}
\caption{VLBI correlator parameters}             
\label{corpars}      
\centering                          
\begin{tabular}{l l l r}        
\hline\hline                 
Source & RA (J2000) & Dec (J2000) & coravg$^1$\\    
\hline                        
YLW\,15 VLA\,2&16h27m26.89s  & $-24^\circ 40'50.9''$   & 4s \\
EC~95        & 18h29m57.92s  & $+01^\circ 12'46.0''$   & 2s \\
HL~Tau       & 04h31m38.432s & $+18^\circ 13'57.485''$ & 2s \\
LDN\,1551$^2$& 04h31m34.158s & $+18^\circ 08'04.760''$ & 2s \\
\hline                                   

\multicolumn{3}{l}{$^1$ correlator averaging time} \\
\multicolumn{3}{l}{$^2$ IRS~5} \\
\end{tabular}
\end{table}

The VLBI data were analyzed as follows. After correcting the antenna parallactic angles recorded in horizontal coordinates and after applying the latest Earth Orientation Parameters, the dispersive delays caused by electrons in the earth's ionosphere were accounted for using Global Positioning System (GPS) models of the electron content of the ionosphere. For amplitude calibration, the $T_{\rm sys}$ and gain curves of the phased VLA (which is recording ratios of antenna to system temperatures) were read, and the flux density scale was set using a calibrator source. Data taken during telescope slewing were flagged. After correcting sampler voltage offsets using autocorrelation total-power spectra, the \textsl{a priori} amplitude calibration was applied based on the $T_{\rm sys}$ and gain tables. In order to remove instrumental residual delays, phase corrections had to be applied, ideally using pulse-cal information inserted into the data during the observations. Since there is no such information in the VLA and Arecibo data, these corrections had to be derived from the fringes on a strong source for baselines involving these telescopes. From them, the residual delay and phase can be computed which can then be taken into account. Finally, global time-dependent phase errors have to be corrected (e.g., phase gradients between the IFs). These are mainly due to inaccuracies in the geometrical delays assumed during correlation, causing phases varying rapidly with time. They were corrected with the fringe-fitting algorithm devised by \citet{sch83}. Before the final application of calibration information, the program sources were phase-referenced to their respective calibrator sources. As a check of global calibration quality, the calibrator sources were referenced to themselves and subsequently imaged. No polarization calibration was attempted because of the very weak flux densities detected. Maps were created without a $uv$ taper, setting the ROBUST parameter in IMAGR to zero.

\begin{table}
\caption{VLA-only results}             
\label{vlatab}      
\centering                          
\begin{tabular}{l l l r r}        
\hline\hline                 
Source & synth. beam & map peak$^1$ & map rms$^1$\\    
\hline                        
YLW~15       & $1\farcs3 \times 0\farcs6$, PA $-10\fdg3$ & 1.48 mJy$^2$ & 0.024 mJy\\
EC~95        & $3\farcs1 \times 2\farcs4$, PA $-38\fdg5$ & 0.55 mJy     & 0.026 mJy\\
HL~Tau       & $0\farcs9 \times 0\farcs3$, PA $-74\fdg2$ & 0.40 mJy     & 0.024 mJy\\
LDN\,1551$^3$ & $0\farcs9 \times 0\farcs3$, PA $-74\fdg2$ & 0.99 mJy$^4$ & 0.025 mJy\\
\hline                                   
\multicolumn{3}{l}{$^1$ peak flux densities and rms}\\
\multicolumn{3}{l}{$^2$ at position of main source (VLA\,1)}\\
\multicolumn{3}{l}{$^3$ IRS~5}\\
\multicolumn{3}{l}{$^4$ northern component}\\
\end{tabular}
\end{table}

\section{Results}
\label{sect.resu}

\subsection{VLA data}

As stated above, we analyzed the VLA-only data taken of each source as part of the VLBI experiments in order to refine the source positions for the correlator. All sources were detected (Table~\ref{vlatab}). The VLA results for the program sources are shown in Fig.~\ref{vlaonly} and summarized in Table~\ref{vlatab}.

\paragraph{YLW\,15}

For YLW15 with its previously analyzed proper motion, it was especially difficult to extract the position of YLW15-VLA2 out of the VLA data where VLA1 and VLA2 are blended. We chose the south-western extension to the main source VLA1, as seen more clearly in an image of our VLA data restored with a circular $0\farcs4$ beam, smaller than the synthesized beam (Fig.~\ref{vlaonlyYLW15}). It remains unclear whether this emission is due to VLA\,2. This position is unexpected given the predictions of \citealp{cur03}---the positions from their Fig.~2 are plotted into our VLA maps in Fig.~\ref{vlaonly} and Fig.~\ref{vlaonlyYLW15}---but the angular resolution of our VLA data is not good enough to decide this issue. However, already from the previous data it becomes clear that VLA\,2 was moving westward relative to VLA\,1, and it is west of these previous positions where we find emission. On the other hand, the predictions are based on only three positions, thus the orbit is not yet fully determined. Given that the observations discussed by \citet[see also \citealp{gir04}]{cur03} were carried out in 1990, 2000, and 2002, the big change seen since then in only about three years---if VLA2 is in the above-mentioned extension---could possibly be explained by the periastron passage of VLA\,2. More observations are needed in order to reliably determine the orbit.

The integrated flux densities of the two components (VLA~1 and 2) as given by \citet{gir04} are 2.18~mJy and 2.14~mJy in 2000 and 2002, respectively; whereas the integrated flux density of VLA~1 is 1.40 mJy and 1.51~mJy for the two epochs. In our case, the integrated flux density for both sources is 2.18~mJy. Assuming therefore no significant long-term variability, the flux density of VLA~2 is 0.67\ldots0.78~mJy.

\paragraph{LDN 1551 IRS 5}

Both components of LDN 1551 IRS 5 are of similar flux density. We found the northern component, of special interest here, to be slightly brighter than the southern component in our VLA data while \citet{rod98} found the opposite in 1996 data. The peak flux density in our data of 0.99~mJy at the position of the northern component compares to a peak flux density of $\sim$0.5~mJy at the position of the southern component in their data (Fig.~2 of \citealp{rod98}). The source morphology in both datasets is similar, and differences in the observed relative source positions are much smaller than the beam size. For these reasons and because of the very elongated beam shape, we refrain from discussing proper motions in this system (cf. \citealp{rod03} and \citealp{lim06}).

\paragraph{HL~Tau}

The peak flux density of HL~Tau, as determined from our VLA-only data is 0.40~mJy which is compatible with the peak flux density of $\sim~0.3$~mJy determined by \citet[Fig.~2]{rod94}. Since their measurement in 1992, the source has shifted eastward by about $0\farcs5$. However, this displacement is along the major axis of the elongated synthesized beam in our data which also prevents a confirmation of the slight east-west elongation of HL~Tau apparent in the 1992 data.

\paragraph{EC 95}

In 1995, EC~95 was observed twice at peak flux densities of $0.64\pm0.02$~mJy (01 May, D array) and $0.62\pm0.02$~mJy (20 Jul, A array), respectively; an observation carried out on 21 Sep 1993 (in CnD array configuration) yields a peak flux density of $0.5\pm0.1$~mJy \citep{smi99}. These flux density is very close to the $0.55\pm0.07$~mJy found in our data, although different from the $1.65\pm0.1$~mJy reported by \citet{eir05} for October 1993.

\subsection{VLBI data}

\begin{figure}
  \centerline{YLW\,15 (VLBI)}
  \vspace{2mm}
  \centerline{\hbox{
      \includegraphics*[width=7.5cm, angle=-90, bb=70 70 530 720]{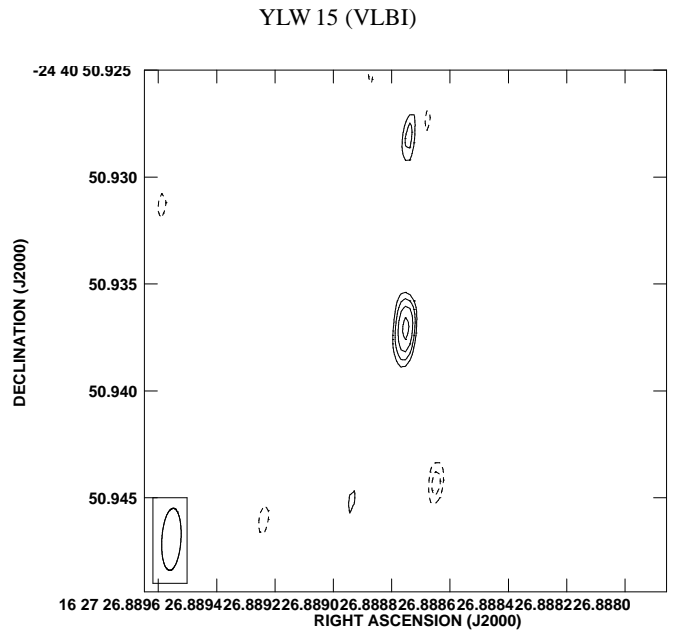}
      }}
\caption{Weak source found in the VLBI data close to the putative position of YLW\,15 YLW\,2 at RA 16h27m26.88875s DEC $-24^\circ40'50.937''$. The source reaches a significance of $\sim9\sigma$. The contour lines delineate multiples of $3\sigma$, increasing by factors of $\sqrt{2}$ as explained for Fig.~\ref{vlaonly}. The $1\sigma$ rms noise level is 15.4~$\mu$Jy.}
\label{vlbiYLW15}
\end{figure}

While all four target sources were detected with the VLA (observing in various configurations, see above), the clean maps produced from the VLBI data contain only one detection. Basic results for each source, including the rms noise level achieved, are listed in Tab.~\ref{vlbitab}. 

In the VLBI data of YLW\,15-VLA\,2, we find a persistently detected weak source (see Fig.~\ref{vlbiYLW15}). In particular, the detection is reproduced when using different cell sizes in image reconstruction and different weighting parameters in the $uv$ plane (ROBUST parameter in AIPS or using a $uv$ taper to reduce the weights of the visibilities with the longest baselines), yielding a significance of $7\ldots9\sigma$. 

The peak flux density of this unresolved source is $0.145\pm 0.016$~mJy, therefore almost 20$\%$ of the flux density of VLA~2 is due to compact emission of $< 0.4 \times 0.1$ AU (FWHM beam size). The flux density is too low for a discussion of detections on single baselines, but the use of all phase-referenced data on all baselines for the full time of the observation allows the detection of the source. Towards VLA\,1, no compact emission was detected. Interestingly, the direction of the outflow towards YLW~15 as seen in CO appears to be quite close to the direction of the line of sight because the integrated line wing emission overlaps considerably \citep{bon96}. This may cause a favourable viewing geometry onto the central source along the jet.

\begin{figure}
  \centerline{YLW\,15 (VLT-ISAAC $J$)}
  \vspace{2mm}
  \centerline{\hbox{
      \includegraphics*[width=\linewidth,bb= 15 20 780 655]{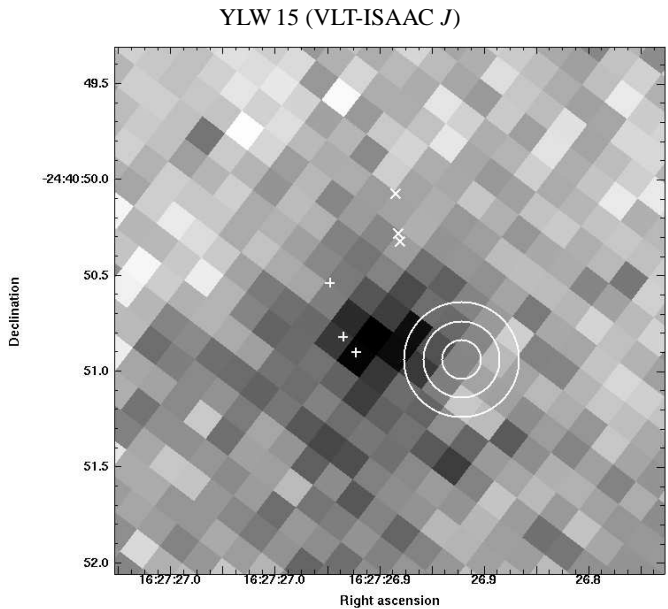}
      }}
\caption{Archival VLT-ISAAC $J$-band data of YLW\,15 (grey-scale image), obtained only 1.5 months after our VLBI run (16 June 2005). Only VLA\,2 is supposedly seen in this band. The VLBI source (marked by a circles with radii of $0\farcs1$, $0\farcs2$, and $0\farcs3$; see text for discussion of astrometric accuracy of the infrared data) is apparently unrelated to the infrared emission. For reference, the positions from previous proper motion analysis are shown as well \citep[][see text]{cur03,gir04}.}
\label{ylw15vltdetail}
\end{figure}

YLW\,15 has been observed in the near-infrared using ISAAC at the ESO Very Large Telescope (VLT) only 1.5 months after our VLBI run (16 June 2005). Since only VLA\,2 is supposedly detected in $J$ band (e.g., \citealp{all02,gir04}), these data can help assess the position of this source independently. Six exposures of 15~sec duration were obtained from the ESO Science Archive (program code 075.C-0561, PI: B. Nisini). They were reduced using the \textsl{jitter} algorithm implemented in the ESO Eclipse package. The accuracy of the astrometry was improved using a fit to the five 2MASS PSC sources in the field of view. With such a low number of reference sources, however, it is difficult to judge the astrometric accuracy, especially since all sources presumably are members of the star-forming region. The general astrometric accuracy of 2MASS is better than $0\farcs1$. A neighbouring source to YLW~15, some $8''$ away is fitted to better than $0\farcs1$ while the remaining 2MASS sources, all quite close to the limits of the field of view though in different directions from YLW~15, are fitted to better than $0\farcs2$. The astrometric precision of the infrared data  is thus estimated to be better than $0\farcs2$, or at least $0\farcs3$. Fig.~\ref{ylw15vltdetail} shows the vicinity of the detected infrared source together with the positions from previous proper motion analysis \citep[][see above]{cur03,gir04} and the position of the weak VLBI source. The infrared emission appears to be unrelated to the VLBI source. 

However, the observed near infrared (1-2 $\mu$m) emission from deeply embedded objects often is not directly due to the protostar, but rather a
combination of stellar light scattered at circumstellar material as well as line emission from shock heated regions in jets and outflows \citep{wei06}. Thus, the position in the $J$-band image might be just light scattered off the dust.

For all sources discussed here, it follows from our observations that the main radio emission observed with the VLA does not come from sub-AU scales around these protostellar sources. This can be explained by radio emission due to an optically thick thermal emission region which is resolved out at VLBI resolution. In fact, while there are some indications of partially non-thermal radio emission in our target sources, there are also clear signs of thermal emission (see source descriptions above). 

\begin{table*}
\caption{VLBI results}             
\label{vlbitab}      
\begin{tabular}{lrrrrr}        
\hline\hline                 
Source & Distance & synth. beam (FWHM) & & map rms$^1$ \\ 
\hline                        
YLW~15       & 130 pc & $2.9~{\rm mas} \times 0.9~{\rm mas}$, PA $-3.6^\circ$ & $0.4~{\rm AU} \times 0.1~{\rm AU}$ & 15.4 $\mu$Jy \\ 
EC~95        & 310 pc & $2.6~{\rm mas} \times 0.7~{\rm mas}$, PA $-7.7^\circ$ & $0.8~{\rm AU} \times 0.2~{\rm AU}$ & 14.3 $\mu$Jy \\ 
HL~Tau       & 140 pc & $1.3~{\rm mas} \times 0.7~{\rm mas}$, PA $-6.3^\circ$ & $0.2~{\rm AU} \times 0.1~{\rm AU}$ & 11.0 $\mu$Jy \\ 
LDN\,1551~IRS\,5 & 140 pc & $1.3~{\rm mas} \times 0.7~{\rm mas}$, PA $-6.1^\circ$ & $0.2~{\rm AU} \times 0.1~{\rm AU}$ & 11.4 $\mu$Jy \\ 
\hline                                   
\end{tabular} \\
$^1$ determined from a Stokes-$I$ clean map with $0\farcs2$ side length ($1024\times 0.2$~mas)\\
\end{table*}

\section{Summary}
\label{sect.summ}

We present the results of a search for non-thermal radio emission on sub-AU scales towards a specially selected sample of protostars showing signs of potentially non-thermal centimetric radio emission. The observations, partly carried out with the High Sensitivity Array, involving most of the largest radio telescopes in the world, at an aggregate data rate of 256\,Mbit\,s$^{-1}$, are the most sensitive yet towards this class of objects. While all sources in this sample are detected in VLA-only observations, we fail to detect compact emission in VLBI observatons except for a persistent detection of compact radio emission towards the putative position of YLW\,15-VLA\,2 with a significance of $7\ldots9\sigma$ depending on the imaging parameters. In this case, the size of the corona can be constrained to $< 0.4 \times 0.1$ AU. Given the apparently fast orbital movement of YLW15\,VLA-2 around VLA-1, more observations will soon enable a much better determination of the orbit. A plausible explanation for the non-detections of the other sources is that thermal free-free emission from ionized regions (e.g., due to stellar winds or jets) is seen in the VLA observations which is resolved out in our VLBI data. 
It seems that the only means to overcome this problem is to look for target sources which have favourable lines of sight allowing for a more direct view of the non-thermal emission in order to reduce the role of the optically thick free-free emission, e.g., by ensuring the clear detection of non-thermal radio emission already in VLA data. Since VLBI observations offer the possibility to study the active coronae and thereby the magnetic fields of protostars at high angular resolution down to $<0.1$~AU, we hope that these observations will be possible for other candidate sources in the future.

\begin{acknowledgements}
We would like to thank the staff of VLBA, VLA, GBT, Effelsberg and Arecibo for carrying out the observations as well as the NRAO data analysts for their support. The National Radio Astronomy Observatory is a facility of the National Science Foundation operated under cooperative agreement by Associated Universities, Inc. Partly based on observations with the 100-m telescope of the MPIfR (Max-Planck-Institut f\"ur Radioastronomie) at Effelsberg. Thanks to Josep Girart for helpful comments. JF acknowledges support by Studienstiftung des deutschen Volkes and the International Max Planck Research School (IMPRS) for Radio and Infrared Astronomy at the Universities of Bonn and Cologne
\end{acknowledgements}

\bibliographystyle{aa} 
\bibliography{bibmaster} 

\end{document}